\documentclass[graybox]{svmult}

\usepackage{type1cm}        
\usepackage{makeidx}         
\usepackage{graphicx}        
\usepackage{multicol}        
\usepackage[bottom]{footmisc}

\usepackage{newtxtext}       %
\usepackage[varvw]{newtxmath}       

\usepackage{hyperref}
\usepackage{bm} 

\makeindex             

\usepackage{subcaption} 

\usepackage{amsmath}
\usepackage{amsfonts}
\usepackage{graphicx}
\usepackage{placeins} 
\usepackage{booktabs}

\begin{document}

\title*{Radiomics-Guided Vision Transformers for Survival Analysis}
\date{}
\author{Qiyuan Shi and Yi Li}
\maketitle

\abstract{Vision Transformers (ViTs) have shown strong empirical performance on high-dimensional medical imaging data, yet their behavior under survival objectives and the interpretability of their attention mechanisms remain poorly understood. Under shallow ViTs, we design controlled experiments showing that token-level attention dynamics can recover outcome-relevant regions and that attention-based thresholding enables effective token pruning, improving both interpretability and predictive performance. We also study pretrained deep ViTs for survival analysis and propose a radiomics-guided hybrid model that integrates pixel-based embeddings with interpretable radiomic features through a multimodal Cox framework and contrastive alignment. Applied to a COVID-19 chest X-ray cohort with a composite ICU admission or mortality endpoint, the proposed approach achieves competitive discrimination while providing clinically meaningful attention maps and feature-group importance.}

\section{Introduction}

The Cox proportional hazards (CoxPH) model has long been a cornerstone of survival analysis in medicine and biostatistics, offering interpretable risk estimation under the proportional hazards assumption. Early extensions incorporated neural networks into the Cox framework, including Cox-nnet \cite{ching2018cox} and DeepSurv \cite{katzman2018deepsurv}, which relaxed linearity and improved predictive performance across genomic and clinical datasets. Other approaches, such as random survival forests, mixture-based Cox models, and DeepHit \cite{lee2018deephit}, further expanded the modeling landscape by accommodating nonlinear effects and competing risks. Despite these advances, most architectures lacked explicit mechanisms to capture complex, high-order dependencies among covariates.

Transformers have recently emerged as a powerful alternative for survival modeling by explicitly modeling feature interactions through attention. \cite{hu2021transformer} introduced the first transformer-based survival model, and SurvTRACE \cite{wang2022survtrace} extended this framework to competing risks with counterfactual learning and interpretable attention. Subsequent work broadened the scope substantially: UniSurv \cite{zhang2025adaptive} proposed a censoring-aware transformer for nonparametric density estimation; SurvFormer \cite{li2025survformer} employed a transformer decoder for imbalanced insurance claims prediction; PAMT \cite{yan2025pathway} combined Vision Transformers (ViTs) on whole-slide pathology with pathway-level genomics for interpretable cancer survival analysis; \cite{cui2025explainable} developed an explainable transformer model for childhood acute lymphoblastic leukemia using SHAP-based attribution; and \cite{wang2025efficient} introduced SRPE-TSN for forecasting adverse events in clinical trials via segmented temporal embeddings.

Together, these studies demonstrate that transformers and ViTs generalize Cox-type models to complex, high-dimensional modalities while offering improved modeling flexibility through attention mechanisms and domain-specific priors. In medical imaging applications, however, interpretability remains a critical concern. Clinical interpretation of images typically relies on two complementary components: identifying relevant spatial regions and understanding which image-derived features drive risk prediction. Any deep learning model intended for clinical use must therefore provide insight into both the regions and the features that influence its predictions. 

For ViTs, the first question is naturally tied to their patch-based representation, in which an image is represented as a sequence of patch tokens and their interactions are modeled through self-attention. In principle, attention weights may reveal which image regions contribute most strongly to a prediction. However, interpreting patch-level importance remains challenging, particularly in high-resolution images where the number of tokens can be large. One approach to understanding patch importance is through token pruning and attention analysis. Token pruning methods selectively remove less informative patches during training or inference, reducing computational cost while retaining predictive performance \cite{ishibashi2025automatic, jaradat2025efficient}. Beyond computational efficiency, these approaches also provide insight into which image regions are most influential for prediction, since pruning decisions are often based on attention scores or related importance measures. Despite a rapidly growing literature on pruning and token selection in ViTs \cite{mao2025prune, jie2025advancing, liu2025evolutionvit, xu2025tafp, lin2025adaptive, zhang2025partitioned}, relatively little work has examined how such mechanisms behave in survival modeling settings.

A second challenge concerns the representation of image features. Most pretrained ViT models rely on pixel-based embeddings derived directly from image patches. While effective for large-scale representation learning, these embeddings do not directly correspond to clinically interpretable descriptors. In medical imaging, structured features such as radiomics descriptors—capturing intensity, texture, and shape patterns—are widely used to characterize imaging phenotypes and provide interpretable summaries of image content. Integrating such structured features with transformer architectures therefore offers a promising avenue for improving interpretability in survival prediction tasks.

To address these questions, we pursue two complementary directions. In Section 3, we use shallow ViTs to study how attention evolves under survival supervision and whether token-level attention can recover outcome-relevant regions. In Section 4, we develop a radiomics-guided deep ViT framework that integrates pixel-based embeddings with interpretable radiomic features within a multimodal Cox model for survival prediction in clinical imaging data. Section 5 concludes with a discussion.

\section{Clinical Interpretation of CXR Images}
Despite the growing use of deep learning for medical image analysis, there is no universally agreed-upon definition of interpretability in medical imaging. In practice, interpretability has been operationalized in several distinct ways. Some studies emphasize patch-level or region-level attention, arguing that spatial localization of salient image areas provides visual explanations aligned with clinical reasoning. Others focus on feature-level importance, including radiomics descriptors or learned latent features, interpreting model coefficients or attribution scores analogously to regression models. In clinical practice, interpretability may also arise from structured scoring systems designed by physicians, which summarize disease severity using predefined anatomical regions and visual criteria. Here we summarize several common perspectives based on the literature (Table~\ref{tab:int}).

\begin{table}[ht]
\centering
\caption{Common perspectives on interpretability in medical image modeling.}
\label{tab:int}
\begin{tabular}{p{4cm} p{6cm} p{3.5cm}}
\toprule
\textbf{Category} & \textbf{Description} & \textbf{Focus} \\
\midrule

Region-Based (e.g., attention localization \cite{chetoui2022explainable})
& Spatial localization of image regions identified by a model.
& Patch row 3, col 5 \\

Feature-Based (e.g., radiomics \cite{prinzi2023explainable})
& Quantified importance or coefficients assigned to extracted features.
& Gray level non-uniformity \\

Manual Score-Based (e.g., RALE \cite{warren2018severity}; Brixia \cite{borghesi2020covid})
& Clinician-designed scoring systems using fixed regions and predefined visual criteria.
& Right lung upper quadrant opacity \\

\bottomrule
\end{tabular}
\end{table}

Although these perspectives are often studied separately, they share two key components: regions of interest (ROIs) and extracted features. Region-based interpretations aim to identify where clinically relevant patterns appear in an image, whereas feature-based interpretations explain predictions through measurable imaging characteristics derived from those regions. In practice, meaningful imaging representations typically require both components, since features must be computed within spatial regions of the image.

In ViTs, images are represented as sequences of patch tokens, where each token corresponds to a local region of the image and its embedding encodes features derived from that region. Consequently, interpretability in ViT-based models can be examined from two complementary perspectives: identifying which patches contribute most strongly to model predictions and understanding which features associated with those patches influence the estimated risk.

These observations motivate the two complementary directions studied in this chapter. We first examine attention dynamics in shallow ViTs to understand how survival supervision influences the selection of outcome-relevant patches, providing a region-level perspective on interpretability. We then study a radiomics-guided deep ViT framework that incorporates radiomic feature representations into survival modeling, enabling feature-level interpretation of risk predictions in clinical imaging data.

\section{Attention Dynamics Under Shallow Survival Transformers}
This section presents synthetic experiments to study how attention evolves under survival supervision in a simplified transformer setting. We use a shallow ViT formulation to enable direct examination of token-level attention dynamics under the Cox objective. Through controlled simulations, we investigate attention concentration, recovery of outcome-relevant tokens, and the effect of attention-based pruning on model performance.

A ViT represents an image as a sequence of tokens rather than raw pixels. The image is first partitioned into non-overlapping patches, which are flattened and mapped to fixed-dimensional vectors through a linear embedding layer. More generally, token embeddings may also be constructed from convolutional backbones such as ResNet \cite{he2016deep} or from handcrafted representations such as radiomic features, depending on the application. In all cases, the resulting image representation is a collection of token vectors that interact globally through self-attention. Introduced for vision tasks by \cite{dosovitskiy2021imageworth16x16words} and building on the original Transformer architecture of \cite{vaswani2017attention}, ViTs have become a standard framework for modeling structured image representations.

Self-attention forms the core of the transformer architecture. In standard matrix notation, it is written as
\[
\operatorname{Attention}(\mathbf{Q},\mathbf{K},\mathbf{V})
=
\operatorname{softmax}\!\left(\frac{\mathbf{Q}\mathbf{K}^\top}{\sqrt{d_k}}\right)\mathbf{V},
\]
where the softmax weights quantify how strongly each token attends to the others, and the factor $\sqrt{d_k}$ rescales the query-key inner products by the key dimension. Each token embedding is linearly projected into query, key, and value representations. In deep ViTs, this mechanism is typically implemented with multi-head attention, layer normalization, feedforward layers, and residual connections. However, when the goal is to study attention learning dynamics directly, these additional architectural components can obscure the role of the attention mechanism itself. For this reason, many theoretical studies of self-attention adopt simplified one- or two-layer transformer architectures to ensure analytical tractability (e.g., \cite{huang2023context}).

Among these, \cite{li2023theoretical} is most closely aligned with our goals and we adopt its notation and settings as a starting point for studying attention dynamics in ViT-based survival models. In parallel, a large body of work has investigated token pruning and selection strategies in ViT. These methods aim to reduce redundancy in the token sequence by removing patches that contribute little to the final prediction. Approaches include gradient-aware pruning \cite{ishibashi2025automatic}, attention-based importance scoring \cite{mao2025prune}, hybrid static--dynamic pruning strategies \cite{jie2025advancing}, and multi-objective pruning frameworks that balance accuracy and computational cost \cite{liu2025evolutionvit}. While these studies are often motivated by computational efficiency, they also provide practical mechanisms for identifying informative tokens and analyzing how attention distributes across image regions.

Motivated by these developments, we examine how attention behaves under survival supervision in a simplified transformer setting. By adopting the shallow architecture used in prior theoretical work, we can directly observe how gradient updates influence attention concentration and whether the model progressively focuses on outcome-relevant tokens. Each observation
$\mathbf{X}_n = [\mathbf{x}_{n1}, \ldots, \mathbf{x}_{nL}] \in \mathbb{R}^{d \times L}$
consists of $L$ tokens, where each token $\mathbf{x}_{n\ell} \in \mathbb{R}^d$ represents a $d$-dimensional embedding derived from image patches or feature regions. Let $S_n \subseteq [L]$ denote the subset of active tokens after sparsification. The model output is defined as
\[
F(\mathbf{X}_n)
=\frac{1}{|S_n|}
\sum_{\ell\in S_n}
\mathbf{a}(\ell)^\top
\operatorname{ReLU}\!\Bigg(
\mathbf{W}_O
\Big[
\sum_{s\in S_n}
\alpha_{s\ell}(\mathbf{X}_n)\,\mathbf{W}_V \mathbf{x}_{ns}
\Big]
\Bigg),
\]
where the attention coefficients are given by
\[
\alpha_{s\ell}(\mathbf{X}_n)
=\frac{\exp\!\big((\mathbf{W}_Q \mathbf{x}_{n\ell})^\top (\mathbf{W}_K \mathbf{x}_{ns})\big)}
{\sum_{j\in S_n}\exp\!\big((\mathbf{W}_Q \mathbf{x}_{n\ell})^\top (\mathbf{W}_K \mathbf{x}_{nj})\big)}.
\]
Here, $\mathbf{W}_Q, \mathbf{W}_K \in \mathbb{R}^{d_k \times d}$, $\mathbf{W}_V \in \mathbb{R}^{d_v \times d}$, and $\mathbf{W}_O \in \mathbb{R}^{m \times d_v}$ denote the query, key, value, and output projection matrices, respectively, and $\mathbf{a}(\ell) \in \mathbb{R}^{m}$ is the linear weight vector associated with token~$\ell$. This single-head formulation follows \cite{li2023theoretical} and omits the $\tfrac{1}{\sqrt{d_k}}$ scaling commonly used in deep multi-head transformers, which simplifies the analysis of gradient dynamics and attention concentration.

\begin{table}[ht]
\centering
\caption{Notation and dimensions for the shallow ViT model.}
\label{tab:vit_notation}
\begin{tabular}{l c l}
\hline
\textbf{Symbol} & \textbf{Dimension} & \textbf{Description} \\
\hline
$\mathbf{X}_n = [\mathbf{x}_{n1}, \ldots, \mathbf{x}_{nL}]$ & $\mathbb{R}^{d \times L}$ & Token matrix for sample $n$ \\
$\mathbf{x}_{n\ell}$ & $\mathbb{R}^{d}$ & $\ell$-th token embedding \\
$S_n$ & -- & Index set of active tokens \\
$\mathbf{W}_Q,\, \mathbf{W}_K$ & $\mathbb{R}^{d_k \times d}$ & Query and key projection matrices \\
$\mathbf{W}_V$ & $\mathbb{R}^{d_v \times d}$ & Value projection matrix \\
$\mathbf{W}_O$ & $\mathbb{R}^{m \times d_v}$ & Output projection matrix \\
$\mathbf{a}(\ell)$ & $\mathbb{R}^{m}$ & Linear weight vector for token $\ell$ \\
$\alpha_{s\ell}(\mathbf{X}_n)$ & scalar & Attention coefficient from token $s$ for $\ell$ \\
$F(\mathbf{X}_n)$ & scalar & Model output (risk score) \\
$d,\, d_k,\, d_v,\, m,\, L$ & -- & Model dimensions and token count \\
\hline
\end{tabular}
\end{table}

This formulation admits a close analogy to modern regression models. The index set $S_n$ plays a role analogous to feature selection, akin to the Lasso \cite{tibshirani1996regression}, while the token-specific vectors $\mathbf{a}(\ell)$ function as regression coefficients. Motivated by this connection, we adopt the shallow ViT as a principled starting point for analyzing attention behavior under the Cox survival objective.

\subsection{Partial likelihood and training dynamics}
To connect the shallow ViT model with survival outcomes, we adopt the Cox proportional hazards framework.
For subject $i$ with event indicator $\delta_i$ and risk score $F(\mathbf{X}_i)$, the negative log-partial likelihood over the full dataset is
\[
L_{N} = -\sum_{i\in [N]} \delta_i \left( F(\mathbf{X}_i) - \log \sum_{j\in R(t_i)} e^{F(\mathbf{X}_j)} \right),
\]
where $R(t_i)$ denotes the risk set at time $t_i$.
This formulation corresponds to full-batch optimization, where all samples are used to update parameters at each iteration.

In practice, when the dataset size $N$ is large or the model is trained with image-based embeddings, gradient updates are performed using stochastic gradient descent (SGD).
In that case, the empirical loss at iteration $b$ is evaluated on a random mini-batch $B_b \subset [N]$:
\[
L_{B} = -\sum_{i\in B_b} \delta_i \left( F(\mathbf{X}_i) - \log \sum_{j\in R(t_i)} e^{F(\mathbf{X}_j)} \right).
\]
Although both $L_N$ and $L_B$ represent the same underlying partial likelihood, the latter is more computationally feasible for large-scale data and induces distinct gradient dynamics compared to full-batch updates.
\cite{zeng2024mini} show that consistency for regression coefficients can still be achieved under mini-batch optimization.
As shown in later sections, both full-batch and mini-batch approaches behave similarly in the training dynamics experiments.

The model parameters $\mathbf{W} = \{\mathbf{W}_Q, \mathbf{W}_K, \mathbf{W}_V, \mathbf{W}_O\}$ are optimized through gradient descent or SGD.
The gradient of the loss with respect to $F(\mathbf{X}_i)$ follows the standard Cox formulation, while its backpropagation through the attention and ReLU layers governs how updates to $\mathbf{W}_Q$ and $\mathbf{W}_K$ influence the evolution of attention concentration across relevant token subsets.

We emphasize that the partial likelihood objective itself is not novel, and the optimization procedure follows the standard deep survival modeling framework (e.g., \cite{wei2021deep}). However, it is well known that the Cox partial likelihood is non-separable due to risk-set coupling, as each gradient term depends on all individuals within the corresponding risk set. In mini-batch SGD, the full risk-set structure is typically approximated \cite{zeng2024mini}, which introduces additional stochasticity beyond standard i.i.d.\ sample assumptions; prior work has studied convergence behavior of deep survival models under such approximations. 

Although our focus is not on asymptotic results, a similar issue related to the risk set arises in extending \cite{li2023theoretical} to the survival setting. Here we briefly introduce the challenge:

In the binary classification case, the mini-batch gradient for any parameter block $\mathbf{W}$ takes the separable form
\[
\frac{\partial L_{\mathrm{Hinge}}}{\partial \mathbf{W}}
=
\frac{1}{B}
\sum_{n\in\mathcal B} (-y_n)\, H(\mathbf{X}_n; \mathbf{W}),
\qquad
y_n \in \{-1,1\},
\]
where $H(\mathbf{X}_n;\mathbf{W})$ denotes a sample-specific quantity that depends only on $\mathbf{X}_n$ and $\mathbf{W}$ and is bounded.
Each summand $(-y_n) H(\mathbf{X}_n;\mathbf{W})$ therefore depends only on the single sample $(\mathbf{X}_n,y_n)$, so the gradient is a sum of independent bounded terms. Under the i.i.d.\ assumption and balanced label distribution, Hoeffding-type concentration inequalities can be applied directly to the mini-batch average to control deviations from the expectation, which is a key step in their inductive analysis.

In contrast, for the Cox partial likelihood, the gradient with respect to any parameter block $\mathbf{W} \in \{\mathbf{W}_Q,\mathbf{W}_K,\mathbf{W}_V,\mathbf{W}_O\}$ can be written as
\[
\frac{\partial L_{\mathrm{Cox}}}{\partial \mathbf{W}}
=
\sum_{n:\delta_n=1} \Bigg(
-
\frac{\partial F(\mathbf{X}_n)}{\partial \mathbf{W}}
+
\sum_{j:\,Y_j \ge Y_n} \pi_{j|n}\,
\frac{\partial F(\mathbf{X}_j)}{\partial \mathbf{W}}
\Bigg),
\qquad
\pi_{j|n} = \frac{\exp\{F(\mathbf{X}_j)\}}{\sum_{k:\,Y_k \ge Y_n} \exp\{F(\mathbf{X}_k)\}}.
\]
Although the bound of $\partial F(\mathbf{X}_j)/\partial \mathbf{W}$ can be investigated, the second (risk-set) sum couples the gradient across many individuals. Consequently, the mini-batch gradient cannot be expressed as an average of independent per-sample contributions, and the Hoeffding-type concentration argument used in \cite{li2023theoretical} does not directly extend to this survival setting.

\subsection{Attention dynamics under survival supervision: designed simulation studies}
\label{sec:exp_dynamics}
As discussed above, the core of the self-attention mechanism in a ViT lies in the four projection matrices that govern how information flows and aggregates across tokens. All other components remain fixed, so the evolution of these matrices determines how attention is allocated during optimization. Extending the shallow-ViT framework of \cite{li2023theoretical} to the Cox partial likelihood setting, we design the following simulations to demonstrate that under survival supervision, the attention weights progressively concentrate on label-relevant tokens that are most predictive of event risk.

We assume each sample $\mathbf{X}_n = [\mathbf{x}_{n1}, \ldots, \mathbf{x}_{nL}] \in \mathbb{R}^{d\times L}$ is composed of $L$ tokens indexed by $S_n = \{1,\ldots,L\}$, where each token arises as a noisy realization of one of several underlying patterns in $\mathbb{R}^d$.
Specifically, there exist $M$ latent prototype vectors $\{\boldsymbol{\mu}_1, \boldsymbol{\mu}_2, \dots, \boldsymbol{\mu}_M\} \subset \mathbb{R}^d$ such that each observed token $\mathbf{x}_{n\ell}$ lies within a small perturbation of one prototype:
\[
\min_{j \in [M]} \|\mathbf{x}_{n\ell} - \boldsymbol{\mu}_j\| \le \tau,
\quad \text{where} \quad
\kappa = \min_{i \neq j}\|\boldsymbol{\mu}_i - \boldsymbol{\mu}_j\|.
\]

We assume that the noise level satisfies $\tau < \kappa/4$.
Among these prototypes, $\boldsymbol{\mu}_1$ and $\boldsymbol{\mu}_2$ correspond to label-relevant patterns, while $\boldsymbol{\mu}_3, \ldots, \boldsymbol{\mu}_M$ represent non-discriminative background patterns.
Tokens generated near $\boldsymbol{\mu}_1$ or $\boldsymbol{\mu}_2$ determine the subject’s risk when they are predominant, whereas instances of these patterns appearing as a minority within other subjects are treated as confusion tokens. This assumption serves as an analytical device and does not restrict the formulation, which allows more general configurations.

For survival outcomes, each individual $n$ is characterized by a subset of label-relevant tokens
$\widetilde S_n^\ast \subseteq S_n$. The hazard-relevant subset $S_n^\ast \subseteq \widetilde S_n^\ast$ is defined as the tokens corresponding to the dominant prototype for subject $n$. Specifically, if $\boldsymbol{\mu}_1$ is dominant, then $\beta_\ell = \beta_{\mathrm{neg}}$ for all $\ell \in S_n^\ast$; if $\boldsymbol{\mu}_2$ is dominant, then $\beta_\ell = \beta_{\mathrm{pos}}$ for all $\ell \in S_n^\ast$, where
\[
\beta_{\mathrm{pos}} > 0,
\qquad
\beta_{\mathrm{neg}} < 0.
\]

The individual hazard function is modeled as
\[
\lambda_n(t)
= \lambda_0(t)\,
\exp\!\Bigg(\frac{1}{100}\sum_{\ell \in S_n^\ast} \beta_\ell\Bigg),
\]
where $\lambda_0(t)$ is the baseline hazard.
Here, positive coefficients ($\beta_{\mathrm{pos}}$) represent harmful contributions, while negative coefficients
($\beta_{\mathrm{neg}}$) represent protective (beneficial) effects.
Minority occurrences of label-relevant patterns are treated as confusion tokens and do not contribute directly to the hazard. The quantities $\alpha^\ast$ and $\alpha^\#$ denote the average fractions of label-relevant and
confusion tokens, respectively, across all samples.
Intuitively, this generative model assumes that tokens cluster around a small number of prototype
features, only a subset of which determine the event hazard.
This provides a structured representation linking the token-level feature space of the ViT to individual-level survival risk.

Now, we are interested in how, for each subject $n$ and token $\ell \in S_n$, the total attention mass assigned to the label-relevant subset $S_n^\ast$ evolves over training:

\begin{equation}
\label{eq:attn_mass_survival}
\sum_{i\in S_n^\ast}
\operatorname{softmax}\!\left(
\mathbf{X}_n^\top \big(\mathbf{W}_K^{(t)}\big)^{\!\top} \mathbf{W}_Q^{(t)}\, \mathbf{x}_{n\ell}
\right)_{i}
=
\frac{\displaystyle \sum_{i\in S_n^\ast}
\exp\!\left(\mathbf{x}_{ni}^{\top} \big(\mathbf{W}_K^{(t)}\big)^{\!\top} \mathbf{W}_Q^{(t)}\, \mathbf{x}_{n\ell}\right)}
{\displaystyle \sum_{r\in S_n}
\exp\!\left(\mathbf{x}_{nr}^{\top} \big(\mathbf{W}_K^{(t)}\big)^{\!\top} \mathbf{W}_Q^{(t)}\, \mathbf{x}_{n\ell}\right)}.
\end{equation}

This formulation captures our expectation that training under the Cox loss will increasingly shift attention toward the subset of tokens most informative for survival outcomes.
This conceptual perspective provides the motivation for the following three experiments, which empirically examine how token-level attention evolves under the Cox objective and whether such concentration behavior emerges in practice.

\subsection*{Simulation 1. Learning dynamics}

We generated synthetic data following the mechanism described in \cite{li2023theoretical} to examine how attention evolves during training.
Each sample $\mathbf{X}_n = [\mathbf{x}_{n1}, \ldots, \mathbf{x}_{nL}]$ is constructed from noisy realizations of prototype patterns, among which a subset corresponds to label-relevant (hazard-related) tokens.
The model is trained using the shallow ViT with the Cox partial likelihood loss.

\begin{figure}[t]
    \centering
    \includegraphics[width=\linewidth]{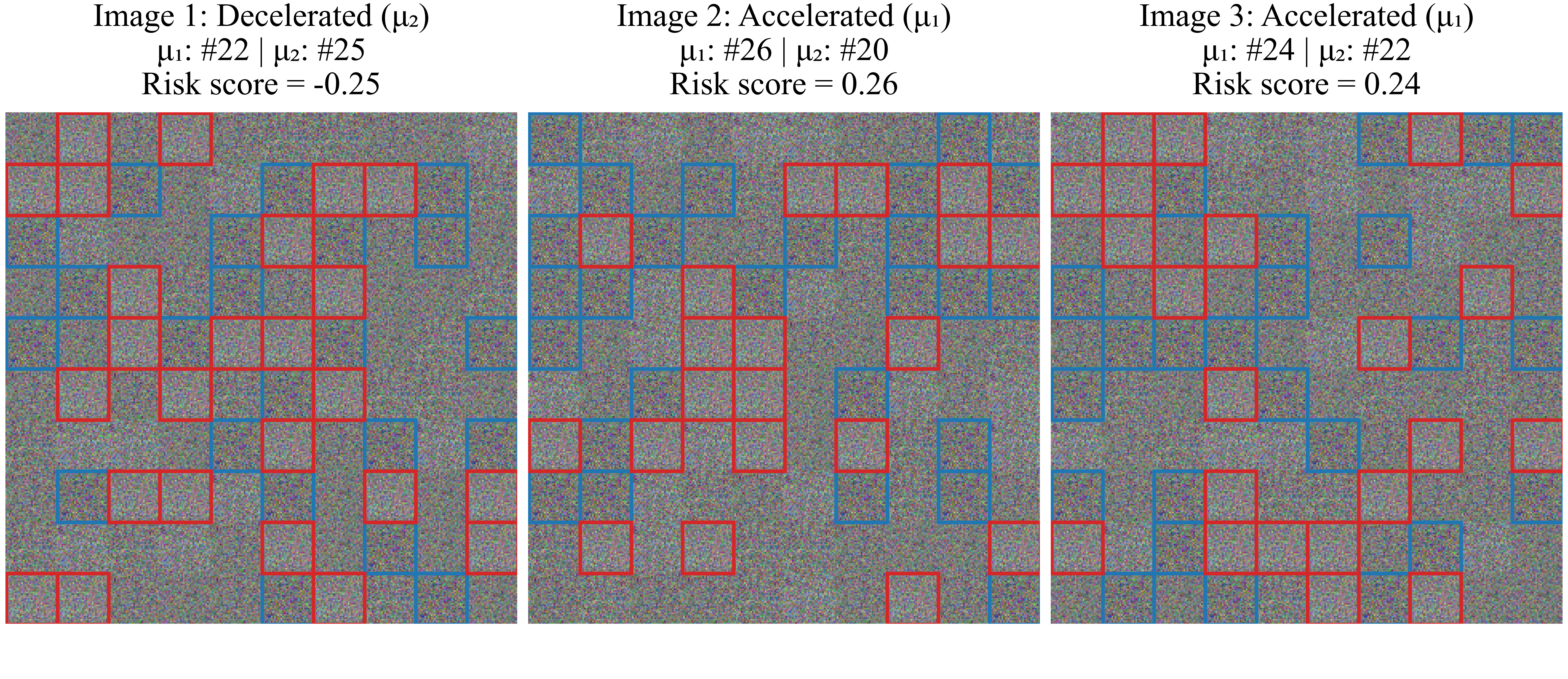}
    \caption{Red boxes correspond to tokens with positive coefficients ($\beta_{\mathrm{pos}}$), and blue boxes indicate tokens with negative coefficients ($\beta_{\mathrm{neg}}$).
This figure demonstrates how, in the ground truth, the risk score is determined by the underlying prototype patterns ($\boldsymbol{\mu}$).
Unlike a majority-vote mechanism, this adaptation to the survival setting allows for a graded, dose–response–like relationship between token patterns and event risk.}
    \label{fig:sim1_concept}
\end{figure}

\begin{figure}[t]
\centering
\includegraphics[
  width=\linewidth,
  height=0.78\textheight,
  keepaspectratio
]{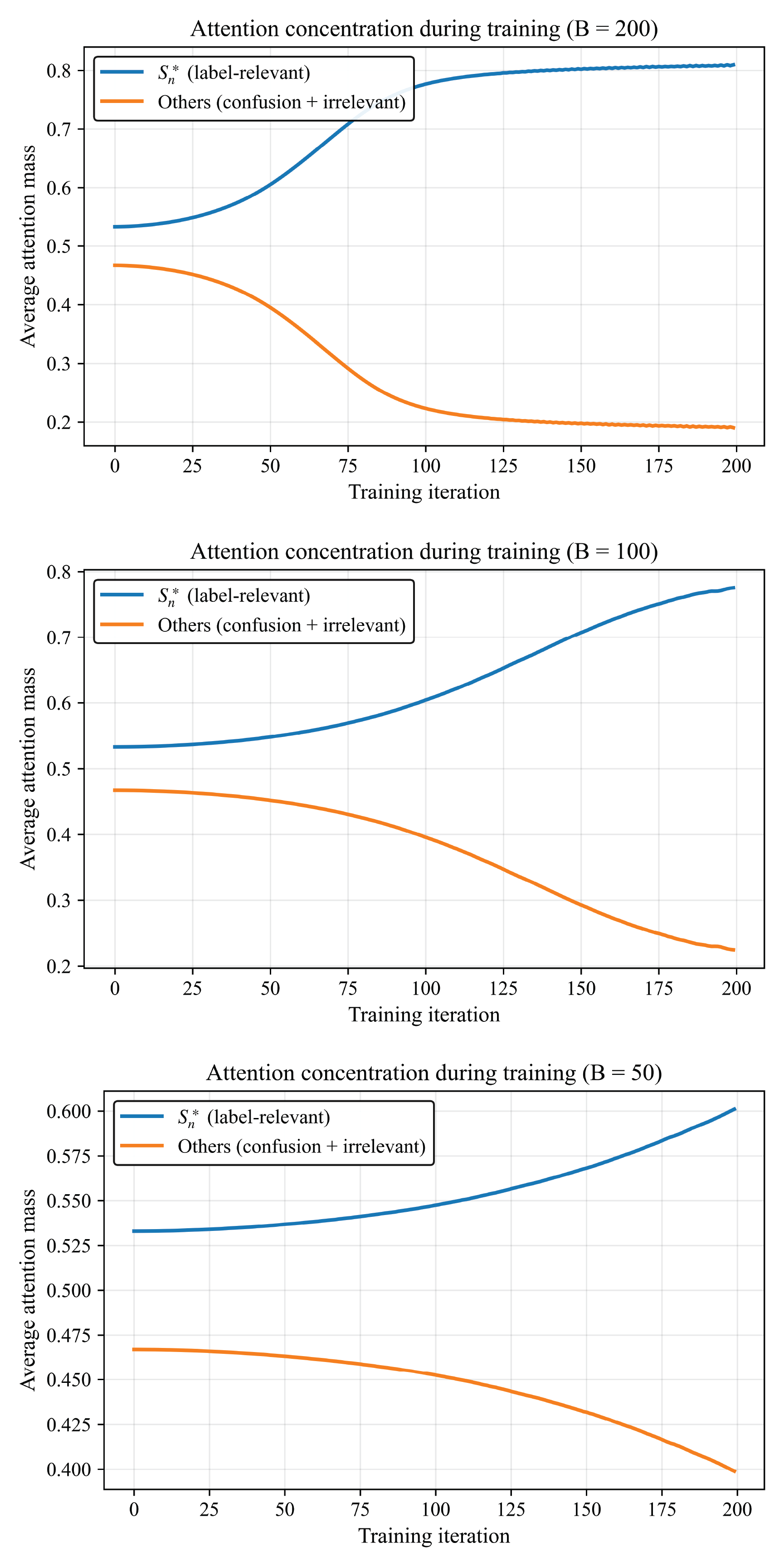}

\caption{Learning dynamics under varying batch sizes $B$ with fixed sample size $N=200$.
Each panel visualizes the evolution of attention concentration across epochs, computed according to Eq.~\eqref{eq:attn_mass_survival}.}
\label{fig:batchsize}
\end{figure}

Figure~\ref{fig:sim1_concept} illustrates how the ground-truth token patterns contribute to the true hazards.
Figure~\ref{fig:batchsize} shows the evolution of attention weights over iterations.
We observe that, consistent with the theoretical prediction, the average attention mass assigned to label-relevant tokens increases steadily during training, while the attention to confusion or irrelevant tokens diminishes.
This validates that, under the Cox objective, the gradient updates to $\mathbf{W}_Q$ and $\mathbf{W}_K$ drive the model toward concentrating attention on label-relevant token subsets.
For both full-batch training and mini-batch training, we observe the same pattern, although with full-batch training the increase is faster.
The sample size is $N = 200$, and all other simulation parameters follow the settings in the data-generating mechanism described earlier.

\FloatBarrier

\subsection*{Simulation 2. Synthetic data for true label-relevant token recovery}

We used the same setup as Simulation~1, except we assumed the label-relevant tokens $S_n^{*}$ were unknown for each observation.
Instead, we computed the average attention each token received from others and, by applying an incremental threshold, manually identified tokens whose attention gains exceeded this threshold as label-relevant.
This experiment examines the practical ability of the shallow ViT to recover the true label-relevant tokens.

\begin{table}[ht]
\centering
\caption{Label-relevant token recovery under a stronger signal.
Fifty repeated datasets are used with $N=200$, $\beta_{\mathrm{pos}}=9.0$, and $\beta_{\mathrm{neg}}=-5.0$.
Fifty percent of tokens are label-relevant and twenty percent are confusion tokens. Results are shown at iteration 200.}
\label{tab:metrics_strong}
\begin{tabular}{c c c c c}
\hline
Threshold & Sensitivity & Specificity & False Positive Rate & False Negative Rate \\
\hline
0.0 & 1.000 ± 0.003 & 0.652 ± 0.066 & 0.348 ± 0.066 & 0.000 ± 0.003 \\
0.2 & 0.974 ± 0.051 & 0.782 ± 0.085 & 0.218 ± 0.085 & 0.026 ± 0.051 \\
0.5 & 0.479 ± 0.191 & 0.938 ± 0.059 & 0.062 ± 0.059 & 0.521 ± 0.191 \\
\hline
\end{tabular}
\end{table}

\begin{table}[ht]
\centering
\caption{Label-relevant token recovery under a weaker signal.
Fifty repeated datasets are used with $N=200$, $\beta_{\mathrm{pos}}=5.0$, and $\beta_{\mathrm{neg}}=-5.0$.
Twenty-five percent of tokens are label-relevant and twenty percent are confusion tokens.
Results are shown at iteration 200.}
\label{tab:metrics_weak}
\begin{tabular}{c c c c c}
\hline
Threshold & Sensitivity & Specificity & False Positive Rate & False Negative Rate \\
\hline
0.0 & 0.846 ± 0.198 & 0.737 ± 0.102 & 0.263 ± 0.102 & 0.154 ± 0.198 \\
0.005 & 0.713 ± 0.242 & 0.826 ± 0.064 & 0.174 ± 0.064 & 0.287 ± 0.242 \\
0.01 & 0.550 ± 0.261 & 0.873 ± 0.067 & 0.127 ± 0.067 & 0.450 ± 0.261 \\
\hline
\end{tabular}
\end{table}

In Tables~\ref{tab:metrics_strong} and~\ref{tab:metrics_weak}, sensitivity is defined as the probability that a truly label-relevant token ($i \in S_n^\ast$) is correctly identified after thresholding its attention gain,
while specificity is the probability that an irrelevant token ($i \notin S_n^\ast$) is correctly excluded.

The stronger the signal, represented by $\alpha^\ast$, the more attention the truly relevant tokens receive.
An ROC analysis can be used to balance sensitivity and specificity, depending on whether interpretability or detection power is prioritized in a given medical context.
Nonetheless, we found that the shallow ViT, despite its simplicity, is already powerful enough to identify outcome-related tokens.
For example, as shown in Table~\ref{tab:metrics_strong}, using a threshold of 0.2 allows 97\% of the true label-relevant tokens to be identified while maintaining a specificity of 0.78.

\subsection*{Simulation 3. Synthetic data for model performance (C-index)}

\begin{figure}[ht]
\centering
\includegraphics[width=0.6\textwidth]{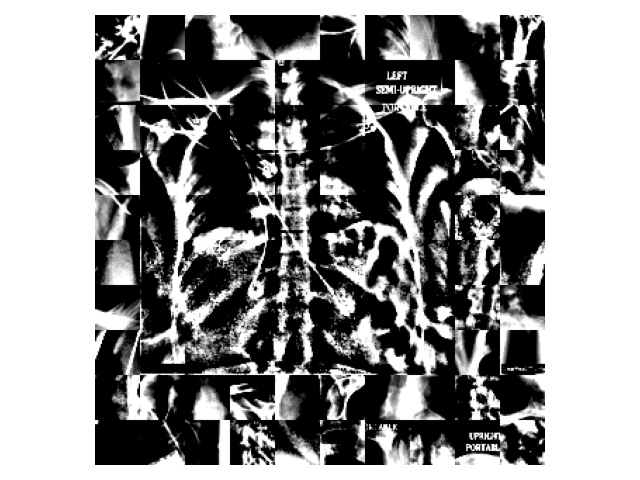}
\caption{Illustration of the embedding expansion process.
The original $7 \times 7$ image tokens are padded with noise to form a $10 \times 10$ token grid.}
\label{fig:fig_embed}
\end{figure}

\begin{table}[ht]
\centering
\label{tab:metrics}
\begin{tabular}{c c   c}
\hline
Model & C-index Test pre-pruning & C-index Test post-pruning \\
\hline
Shallow ViT & 0.67  & 0.70  \\

\hline
\end{tabular}
\caption{This table shows that removal of label-irrelevant tokens or confusion tokens increases overall test performance.}
\end{table}

We evaluated how removing label-irrelevant tokens affects survival prediction performance.
We generated synthetic images by embedding real CXR patches within noisy backgrounds to simulate a mixture of informative and non-informative regions (Figure~\ref{fig:fig_embed}).
A shallow ViT model with Cox loss was trained under two conditions: using all tokens and using only label-relevant tokens identified through attention-based thresholding.
Across replications, pruning confusion tokens consistently improved generalization, with test C-index increasing from approximately 0.67 to 0.70.
This experiment demonstrates that selective token pruning can recover the true discriminative structure and enhance model performance, supporting the theoretical expectation that attention concentration on relevant regions yields better survival prediction.

\section{Radiomics-guided Deep ViTs for Survival Outcomes}

This section focuses on feature-level modeling for survival prediction using radiomics-guided ViT architectures. In contrast to the patch-level attention dynamics studied in controlled shallow settings, we consider real-data survival analysis where feature relevance is not known a priori and must be learned from data. We develop a ViT–Cox framework that integrates pixel-based representations from deep pretrained transformers with radiomics-based descriptors, enabling a unified comparison of pixel, radiomics, and hybrid embeddings. Using COVID-19 chest X-ray data, we evaluate predictive performance and examine how radiomics-guided integration influences model behavior. To enhance interpretability, we analyze attention-based region highlighting and quantify feature-group contributions through coefficient-based importance measures, providing clinically meaningful insights into model predictions.

Radiomics features \cite{lambin2012radiomics,aerts2014decoding}—quantifying intensity, texture, and shape statistics—offer explicit, human-understandable characterizations of images and thus serve as a natural reference to assess whether ViT attention focuses on medically meaningful regions or patterns. These methods are widely applied in medical image analysis. However, a gap remains between such theoretical formulations and real-world ViTs for two main reasons. First, large pretrained models, such as ViTs introduced by \cite{dosovitskiy2021imageworth16x16words} and implemented in the \texttt{timm} library~\cite{Wightman_PyTorch_Image_Models}, are trained purely on pixel intensities and therefore capture broad visual patterns rather than radiomics-style descriptors that explicitly encode intensity, texture, or shape. Second, modern ViTs comprise multiple stacked transformer blocks with multi-head self-attention and feedforward sublayers that iteratively refine patch-level representations, making it nontrivial to directly extend findings from the shallow, two-layer architectures studied in earlier sections.

In fact, several recent studies have begun addressing this gap. At a high level, this represents a broader multimodality challenge that has been extensively explored in works such as CLIP \cite{radford2021learningtransferablevisualmodels} and related frameworks. More specifically, recent research has proposed various fusion strategies to either combine pixel-based and radiomics features or use radiomics information to guide the learning of image-based representations \cite{han2022radiomics,yang2022deep,yang2025embedding}. Although these studies highlight promising directions, it remains unclear how well such multimodal principles translate to survival prediction tasks or whether radiomics and pixel-based representations contribute complementary information. To investigate this, we first establish a baseline full ViT model that serves as a unified framework for both image- and radiomics-based inputs, allowing direct comparison between modalities while isolating the effect of architecture depth and representation type.

In the baseline full ViT framework, we retain the core architectural components of modern transformers---multi-head self-attention, the \texttt{[CLS]} token, and all 12 stacked transformer blocks.
Formally, the prediction for subject \(n\) is based on the final \texttt{[CLS]} token representation:
\[
F_{\text{full}}(X_n)
= \boldsymbol{\beta}^{\top}\operatorname{ReLU}(\text{CLS}_n),
\]
where $\boldsymbol{\beta} \in \mathbb{R}^d$ is the learned linear predictor and $\text{CLS}_n$ is the global embedding from the final transformer layer.
This configuration corresponds to the standard pixel-based ViT, in which all patch-level information is integrated into a single global representation.

Inspired by recent contrastive multimodal approaches such as CLIP and Radiomics-guided fusion models, we extend the full ViT to a dual-branch architecture.
Each subject \(n\) is represented by two latent embeddings: the pixel-derived representation \(\text{CLS}_n\) from the image encoder and the radiomics token vector \(\mathbf{r}_n\) from the radiomics encoder.
Each branch produces a modality-specific scalar risk through a linear projection:
\begin{equation}
r_{n}^{(\text{img})}
= \boldsymbol{\beta}^{\top}\phi(\text{CLS}_{n}),
\qquad
r_{n}^{(\text{rad})}
= \boldsymbol{\gamma}^{\top}\phi(\mathbf{r}_{n}),
\label{eq:dual_rep}
\end{equation}
where $\phi$ denotes an activation function (e.g., ReLU) and $\boldsymbol{\beta}, \boldsymbol{\gamma} \in \mathbb{R}^d$ are the modality-specific coefficients.
The overall mixed risk combines these components as
\begin{equation}
F(X_n)
= (1 - \alpha)\,r_{n}^{(\text{img})}
+ \alpha\,r_{n}^{(\text{rad})},
\qquad
\alpha \in [0,1],
\label{eq:mix_risk}
\end{equation}
allowing flexible control of the radiomics contribution. 
The key point of the mixed risk \eqref{eq:mix_risk} is not the specific functional form or aggregation choice, but the fact that the risk incorporates both outcome-relevant signals and representation-level information. The formulation used is a concrete instance that makes this coupling explicit and estimable. We could consider different functional forms or aggregation strategies, but it is unlikely that they would change the underlying principle. Their impact is secondary and operates within the same framework, affecting finite-sample behavior rather than altering the core mechanism. The results should therefore be interpreted as demonstrating the benefit of integrating representation and outcome information, rather than relying on any particular choice of linearity or aggregation.

The model is trained by minimizing the negative Breslow-style partial log-likelihood:
\[
L_{\text{Cox}}
= - \sum_{i=1}^{N} \delta_i
\left(
F(X_i) -
\log \sum_{j \in R(t_i)} e^{F(X_j)}
\right),
\]
where $\delta_i$ is the event indicator and $R(t_i)$ denotes the risk set at time $t_i$.

To align modalities, we include NT-Xent\cite{pmlr-v119-chen20j}, a cross-view contrastive loss based on the normalized temperature-scaled cross-entropy:
\[
L_{\text{CL}}
= -\frac{1}{N}\sum_{n=1}^{N}
\log
\frac{
\exp\!\big(\mathrm{sim}(\mathbf{z}_n^{(\text{img})},\, \mathbf{z}_n^{(\text{rad})}) / \tau\big)
}{
\sum_{m=1}^{N}
\exp\!\big(\mathrm{sim}(\mathbf{z}_n^{(\text{img})},\, \mathbf{z}_m^{(\text{rad})}) / \tau\big)
},
\]
where $\mathbf{z}_n^{(\text{img})}$ and $\mathbf{z}_n^{(\text{rad})}$ are projection-head embeddings, $\mathrm{sim}(\cdot,\cdot)$ is cosine similarity, and $\tau$ is the temperature parameter.

The final objective jointly optimizes both terms:
\[
L_{\text{total}}
= (1 - \lambda_{\text{loss}})\,L_{\text{Cox}}
+ \lambda_{\text{loss}}\,L_{\text{CL}},
\qquad
\lambda_{\text{loss}} \in [0,1],
\]
balancing survival discrimination and cross-modal consistency.

\begin{figure}[t]
    \centering
    \begin{tabular}{cc}
        \includegraphics[width=0.94\linewidth]{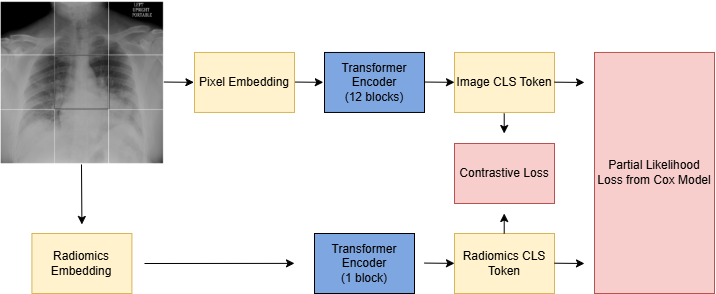} &

    \end{tabular}
    \caption{Overview of the proposed model architecture.}
    \label{fig:fig_arch}
\end{figure}

This hybrid framework is illustrated in Figure~\ref{fig:fig_arch}.
Unlike \cite{han2022radiomics}, the radiomics embeddings were computed per patch in advance, eliminating the need for repeated feature extraction during training. See Table~\ref{tab:cindex_results} for details.

The per-patch region of interest (ROI) extraction also enables direct interpretability.
Another advantage is that the two encoders interact indirectly through the contrastive loss, preserving the pretrained pixel-based ViT’s affinity for pixel embeddings.
Finally, the radiomics encoder is essentially the shallow ViT discussed in an earlier section, which, although not further investigated here, provides a potential basis for future analysis of training dynamics.

\subsection{Real-data application: COVID-19 chest X-ray (Precision Health, University of Michigan)}

We used the University of Michigan Precision Health COVID-19 chest X-ray (CXR) cohort and defined a composite endpoint of ICU admission or death within 28 days of the index imaging. Patients without an event by day 28 were administratively censored at 28 days. We trained three pretrained \texttt{timm} 12-block ViT models and compared them with two radiomics-based approaches:
(i) \texttt{ViT-B/16-384} (patch size 16, input 384),
(ii) \texttt{ViT-B/16-224} (patch size 16, input 224),
(iii) \texttt{ViT-B/32-384} (patch size 32, input 384),
(iv) a \textbf{Radiomics-based} Cox model using handcrafted CXR features, and
(v) a \textbf{Radiomics-guided hybrid ViT} model integrating pixel and radiomics embeddings.

All transformer embeddings had a dimensionality of 768. For the radiomics model, features were extracted using the \texttt{PyRadiomics}\cite{van2017computational} package from 144 non-overlapping $32\times32$ patches per image, yielding approximately 1,218 features per patch. After removing 487 zero-variance features, the remaining features were standardized. The six feature classes (First-order, GLCM, GLRLM, GLSZM, NGTDM, and GLDM) were distributed across the twelve attention heads.

For the Radiomics-guided hybrid ViT, we used the \texttt{ViT-B/32-384} configuration (144 patches). Radiomics features were extracted in the same manner as above, except we divided the features into twelve groups to balance the number of features in each group, ensuring that each group contains features from only one class. Each combination of patch and feature group was then treated as one radiomics token, resulting in $144 \times 12$ tokens aligned with the ViT’s attention structure. The values of $\alpha$ and $\lambda_{\text{loss}}$ were both set to 0.5.

All models were trained using identical five-fold cross-validation with the Adam optimizer ($\text{lr}=10^{-5}$) and batch size 200. The primary evaluation metric was the C-index, reported as mean $\pm$ SD across folds.
\begin{table}[ht]
\centering
\caption{C-index (mean $\pm$ SD across folds) at the epoch with maximum across-fold mean ($\leq 50$ epochs).}
\begin{tabular}{lcccc}
\toprule
\textbf{Model} & \textbf{Train C-index} & \textbf{Test C-index} & \textbf{5 epochs (sec)} & \textbf{10 epochs (sec)} \\
\midrule
Pixels (\texttt{ViT-B/32-384}) & $0.807 \pm 0.006$ & $0.753 \pm 0.011$ & $74.0 \pm 3.3$ & $148.6 \pm 5.0$ \\
Pixels (\texttt{ViT-B/16-384}) & $0.807 \pm 0.004$ & $0.758 \pm 0.011$ & $300.9 \pm 7.5$ & $604.1 \pm 7.4$ \\
Pixels (\texttt{ViT-B/16-224}) & $0.811 \pm 0.005$ & $0.748 \pm 0.023$ & $90.5 \pm 1.3$ & $181.7 \pm 1.6$ \\
Radiomics & $0.727 \pm 0.009$ & $0.706 \pm 0.028$ & $48.5 \pm 2.7$ & $98.5 \pm 3.8$ \\
Radiomics-guided hybrid ViT & $\mathbf{0.992} \pm 0.001$ & $\mathbf{0.756} \pm 0.017$ & $\mathbf{107.3} \pm 1.2$ & $214.6 \pm 2.4$ \\
\bottomrule
\end{tabular}
\label{tab:cindex_results}
\end{table}

\begin{figure}[t!]
\centering
\begin{subfigure}[t]{0.48\linewidth}
    \centering
    \includegraphics[width=\linewidth]{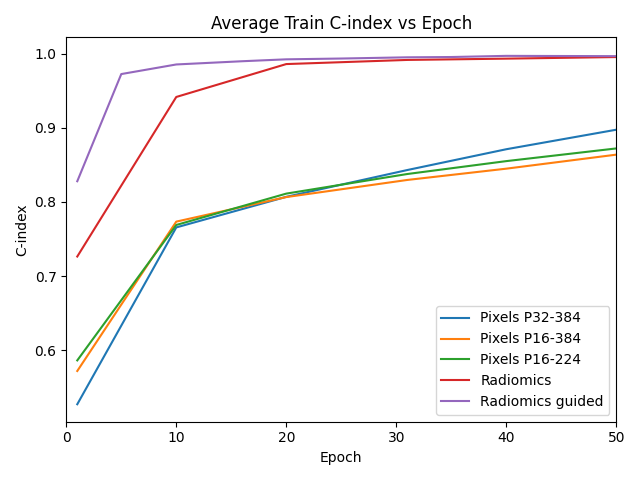}
    \caption{Training}
    \label{fig:vit_auc}
\end{subfigure}
\hfill
\begin{subfigure}[t]{0.48\linewidth}
    \centering
    \includegraphics[width=\linewidth]{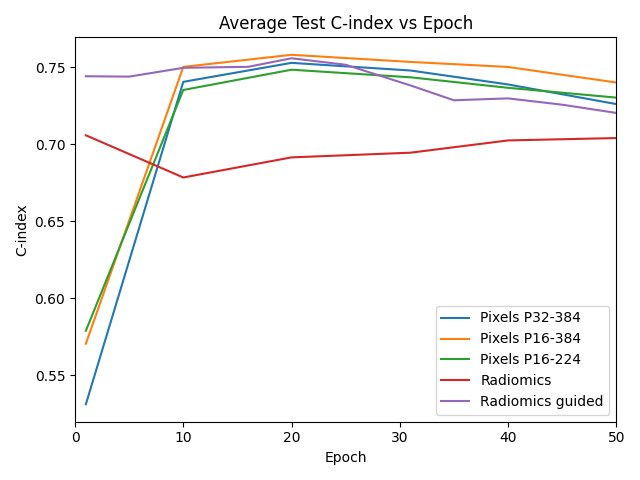}
    \caption{Test}
    \label{fig:vit_calibration}
\end{subfigure}
\caption{Average performance of 5 folds.}
\label{fig:fig_real_training_test}
\end{figure}

Across all configurations, both pixel-based and radiomics-based ViTs achieved competitive survival prediction performance.
However, the naïve use of radiomics features with pixel-based ViTs yielded the lowest accuracy, likely because the pretrained self-attention weights were optimized for image pixel representations rather than radiomics inputs. The weights did not directly capture radiomics patterns meaningfully.

Among the pixel-based models, \textbf{ViT-B/16-384} achieved the best performance, as expected given its higher resolution and finer patch division.
Notably, when using \textbf{ViT-B/32-384} as the backbone, the Radiomics-guided hybrid ViT improved performance, making it nearly comparable to \textbf{ViT-B/16-384}.
Moreover, both the training and testing performance of the hybrid model reached their maxima faster than those of purely pixel-based models, suggesting that radiomics guidance facilitates faster convergence during training (see Table~\ref{tab:cindex_results}). Although the hybrid model is not the fastest per epoch, it reaches near-saturated training performance within the first 5–10 epochs, already achieving approximately 95\% of its peak C-index. In contrast, the purely pixel-based ViTs require substantially more epochs to reach comparable performance levels.

Notably, all ViT models reached peak performance at around 20 epochs, indicating potential overfitting with longer training.
This suggests that an early stopping strategy could be beneficial.

To examine the effect of tuning parameters, we consider a small set of representative $(\alpha, \lambda_{\text{loss}}, \tau)$ values rather than performing exhaustive hyperparameter tuning. As shown in Table~\ref{tab:alpha_lambda_tau_results}, performance was most sensitive to $\lambda_{\text{loss}}$, with smaller values generally yielding better test C-index, whereas moderate changes in $\alpha$ and $\tau$ produced comparatively smaller differences.

\begin{table}[ht]
\centering
\caption{Test C-index (mean $\pm$ SD across folds) for selected $(\alpha,\lambda_{\text{loss}},\tau)$ settings.}
\label{tab:alpha_lambda_tau_results}
\begin{tabular}{ccc c @{\hspace{1.2cm}} ccc c}
\toprule
$\alpha$ & $\lambda_{\text{loss}}$ & $\tau$ & \textbf{Test C-index}
& $\alpha$ & $\lambda_{\text{loss}}$ & $\tau$ & \textbf{Test C-index} \\
\midrule
0.00 & 0.10 & 0.07 & $0.750 \pm 0.018$
& 0.90 & 0.90 & 0.07 & $0.737 \pm 0.013$ \\
0.50 & 0.10 & 0.07 & $0.751 \pm 0.020$
& 0.50 & 0.90 & 0.05 & $0.746 \pm 0.015$ \\
0.90 & 0.10 & 0.07 & $0.751 \pm 0.017$
& 0.50 & 0.90 & 0.20 & $0.743 \pm 0.015$ \\
0.00 & 0.90 & 0.07 & $0.741 \pm 0.012$
& 0.00 & 0.97 & 0.07 & $0.716 \pm 0.027$ \\
0.50 & 0.90 & 0.07 & $0.747 \pm 0.014$
& 0.50 & 0.97 & 0.07 & $0.728 \pm 0.018$ \\
\bottomrule
\end{tabular}
\end{table}

\subsection{Patch inference and visualization: ViT and radiomics}
\begin{figure}[t]
    \centering
    \begin{tabular}{cc}
        \includegraphics[width=0.47\linewidth]{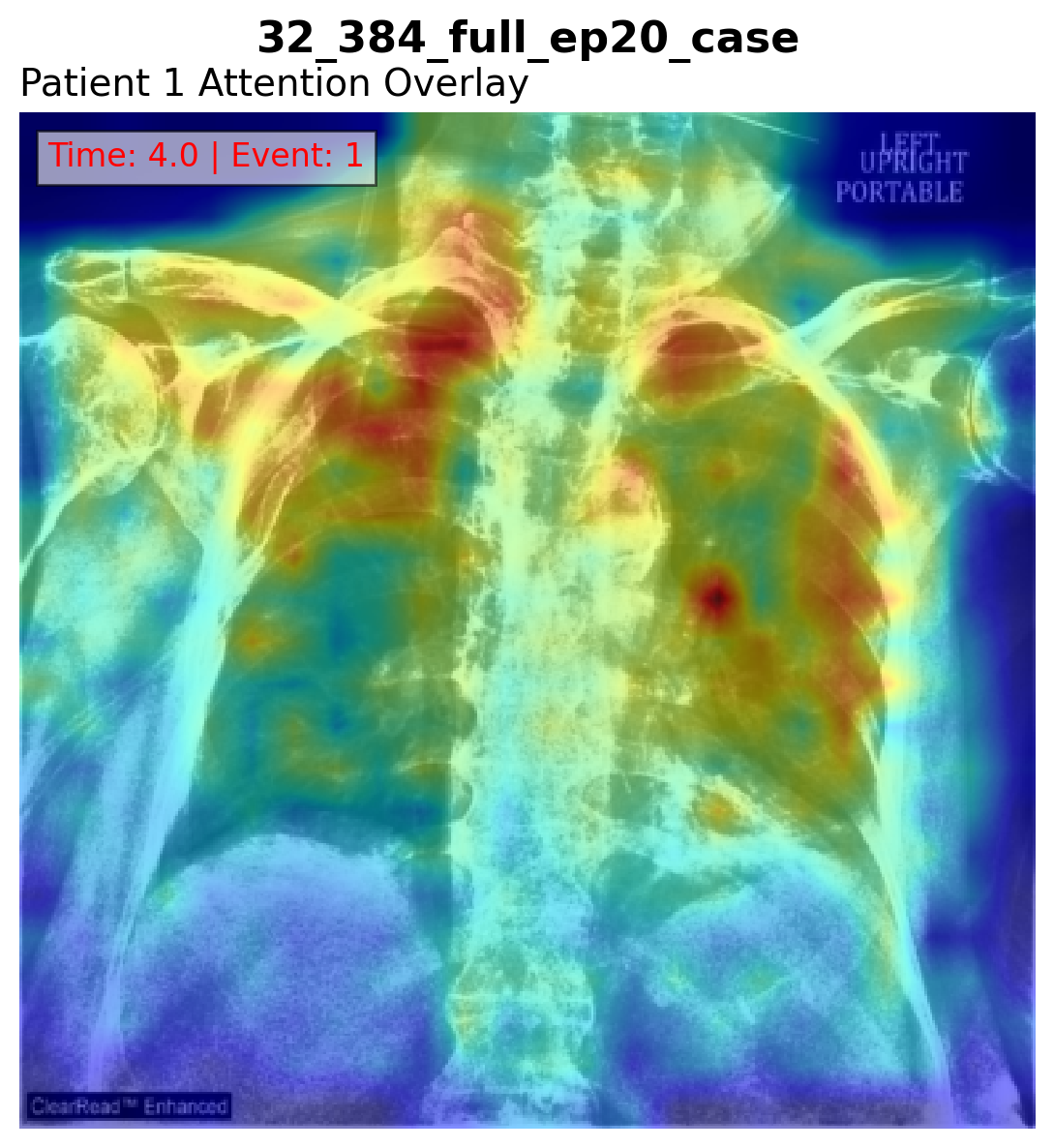} &
        \includegraphics[width=0.47\linewidth]{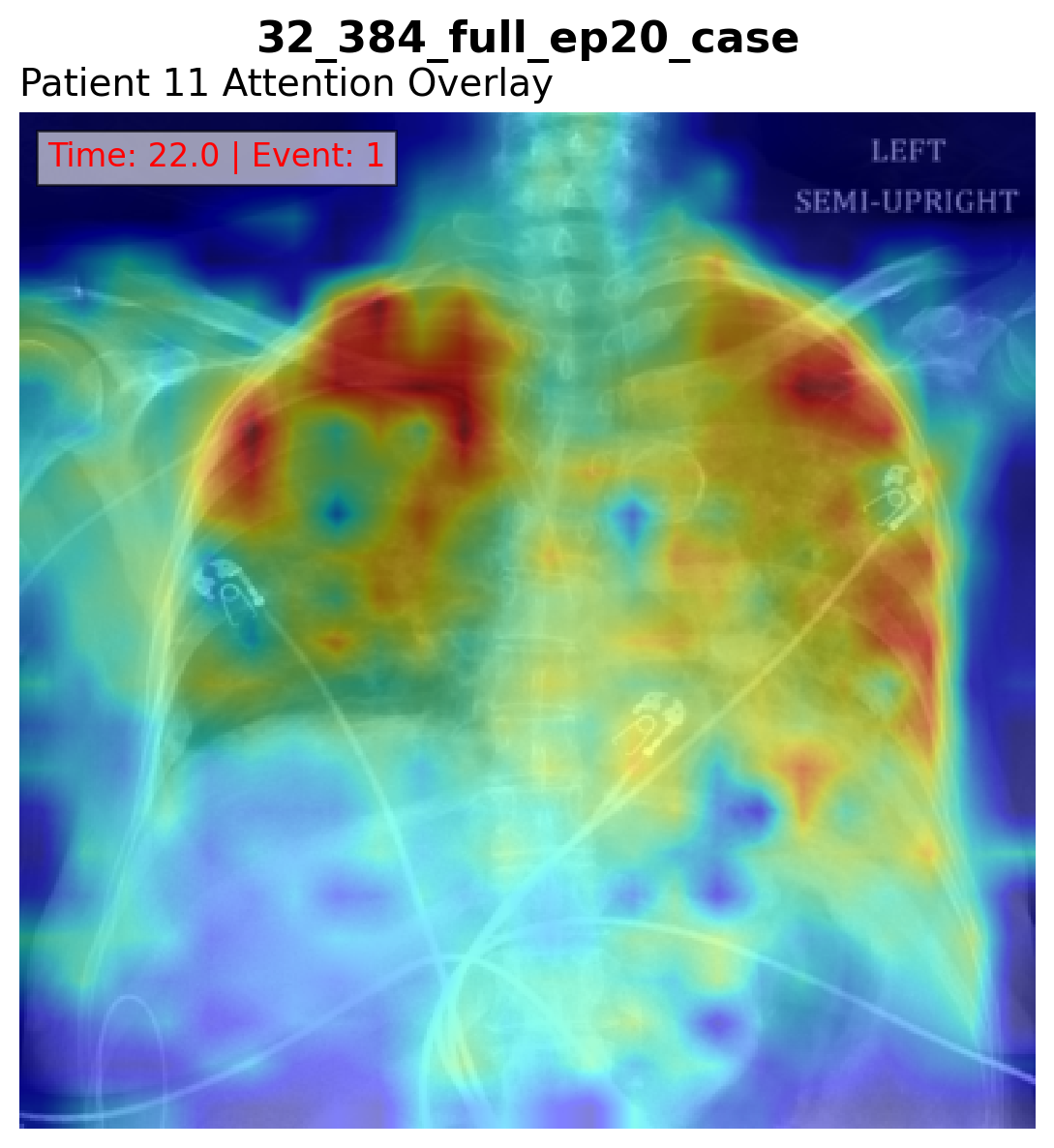} \\[-1mm]

        \includegraphics[width=0.47\linewidth]{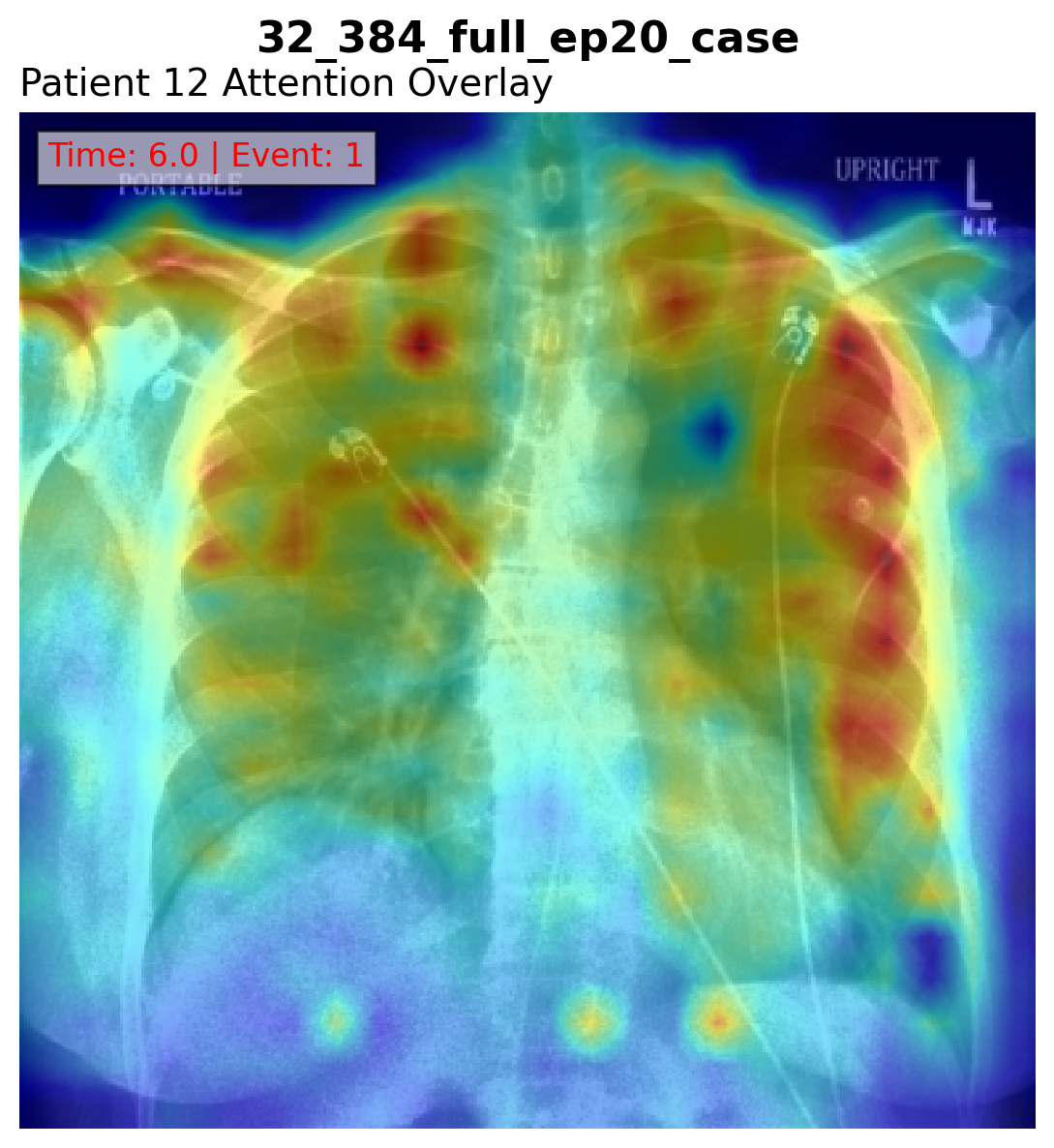} &
        \includegraphics[width=0.47\linewidth]{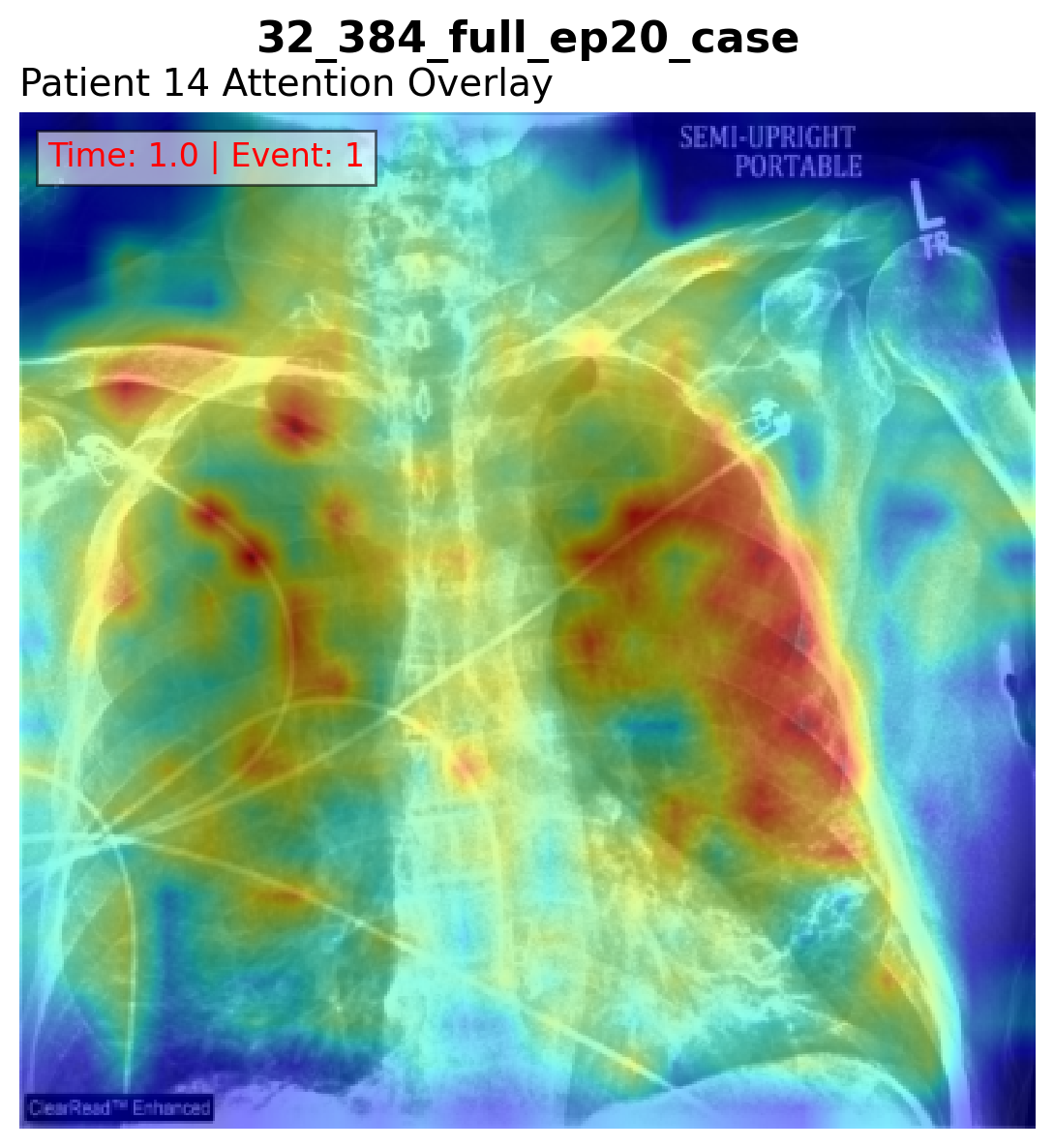} \\[-1mm]

    \end{tabular}
    \caption{An example of CXR for cases from \textbf{ViT-B/32-384}}
    \label{fig:cxr_pix_case}
\end{figure}

\begin{figure}[t!]
\centering
\begin{subfigure}[t]{0.48\linewidth}
    \centering
    \includegraphics[width=\linewidth]{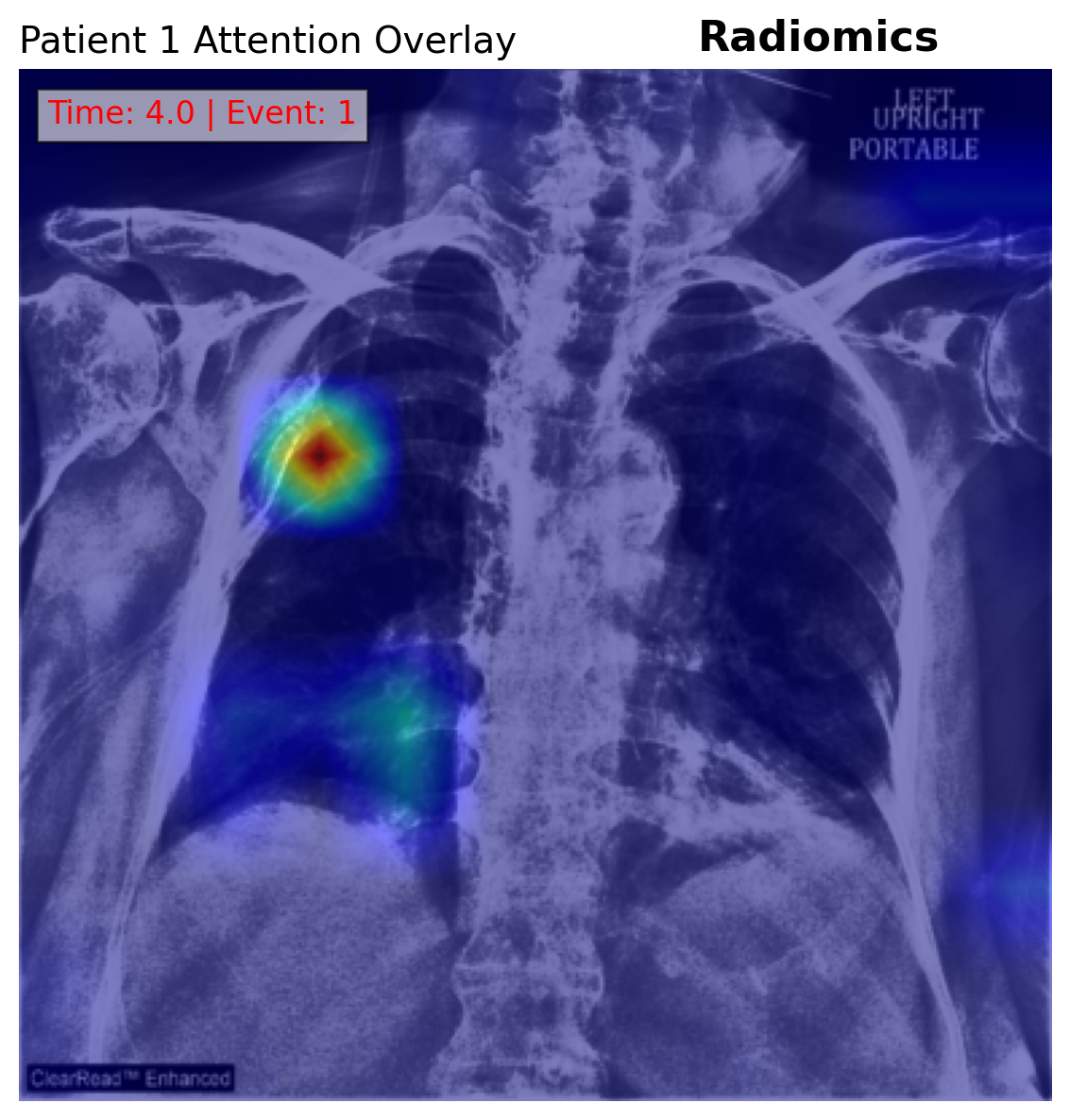}
\end{subfigure}
\hfill
\begin{subfigure}[t]{0.48\linewidth}
    \centering
    \includegraphics[width=\linewidth]{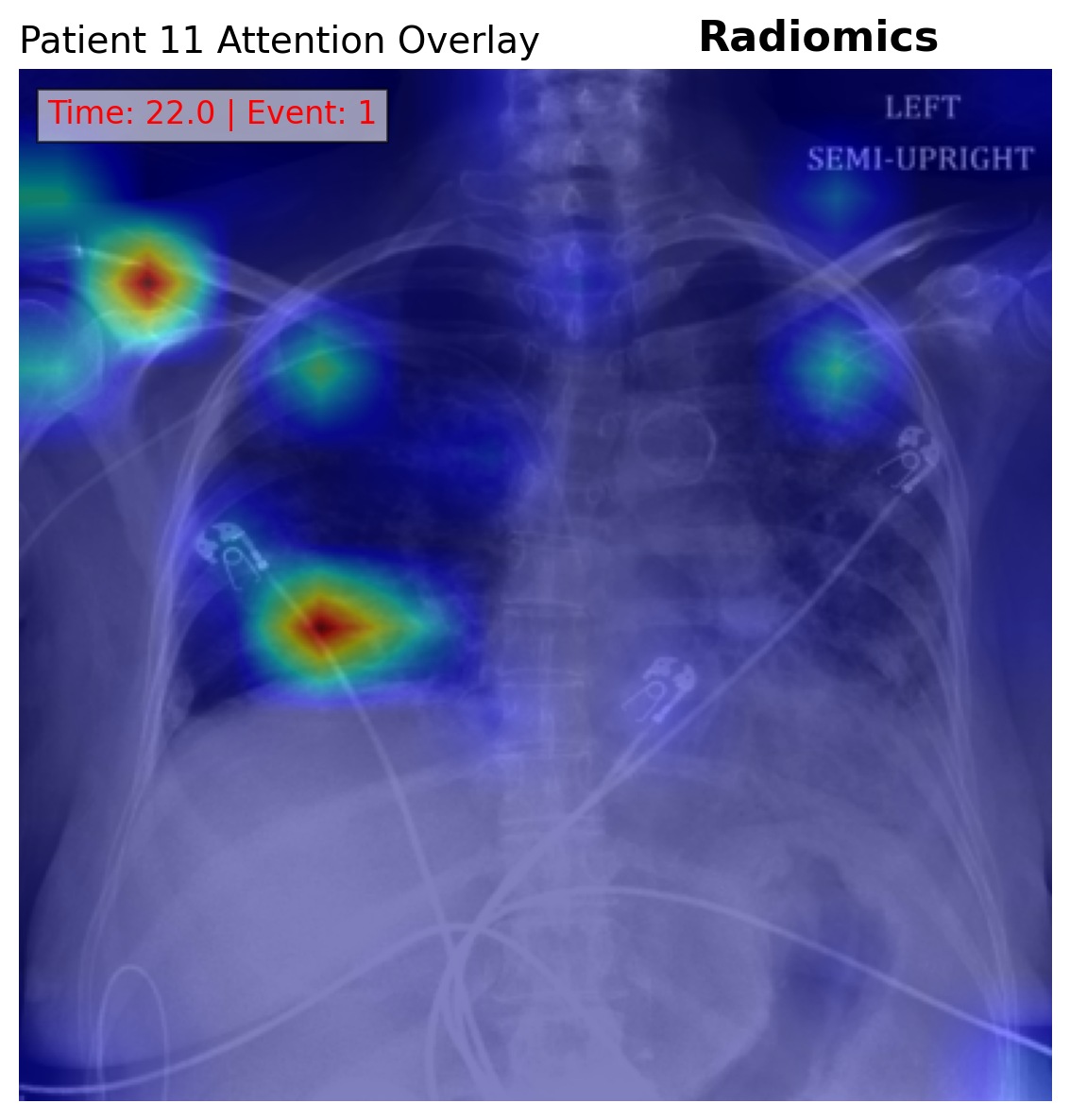}
\end{subfigure}
\caption{An example of CXR for cases from Radiomics model}
\label{fig:cxr_rad_case}
\end{figure}

Figures~\ref{fig:cxr_pix_case} and \ref{fig:cxr_rad_case} illustrate how different model types highlight important regions of the CXRs.
The attention maps of the full ViT models are defined by the attention weights received by the \texttt{[CLS]} token.
As shown, the standard ViT models primarily focus on the lung fields, which is consistent with prior studies \cite{chetoui2022explainable}.
There are shared regions of high attention across patients—particularly the left lung area above the cardiac silhouette—while finer attention patterns vary by individual.
For instance, in Patient~11, where the right hemidiaphragm is elevated and the right lung volume appears reduced, the model assigns greater attention to the right lung, reflecting potential pathological changes.

In comparison, the attention heatmaps from the radiomics-based model are more difficult to interpret.
Only a small number of patches receive high attention, suggesting that the model focuses on limited regions or feature groups that may not have clear spatial correspondence in the image domain.

\FloatBarrier

\subsection{Patch inference and visualization: Radiomics-guided ViT}
\begin{figure}[ht]
\centering
\includegraphics[
    width=\linewidth,
    height=0.8\textheight,
    keepaspectratio
]{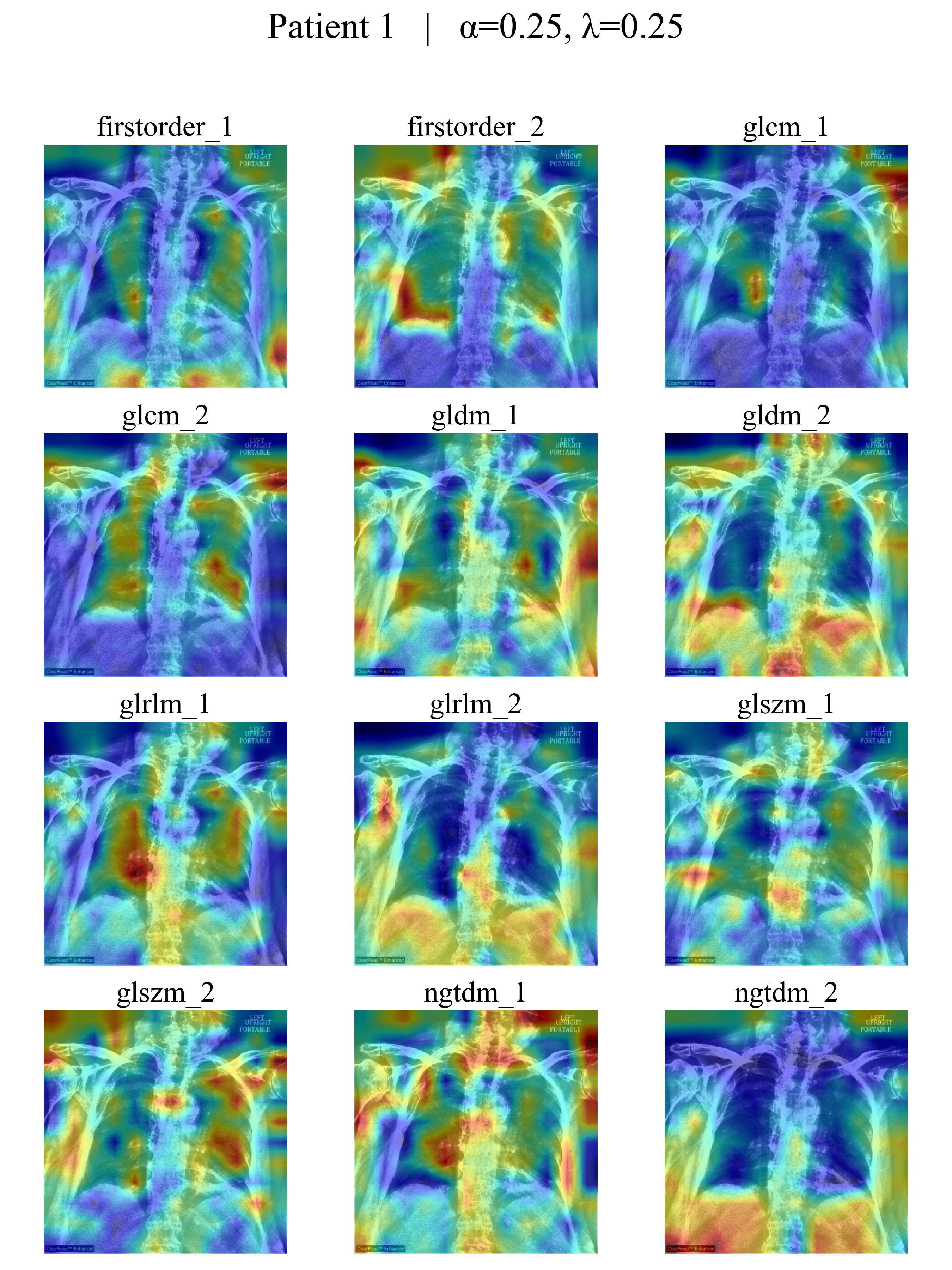}
\caption{CXR with highlighted regions guided by radiomics features (Patient 1).}
    \label{fig:fig_guide_case_1}
\end{figure}

\begin{figure}[ht]
\centering
\includegraphics[
    width=\linewidth,
    height=0.8\textheight,
    keepaspectratio
]{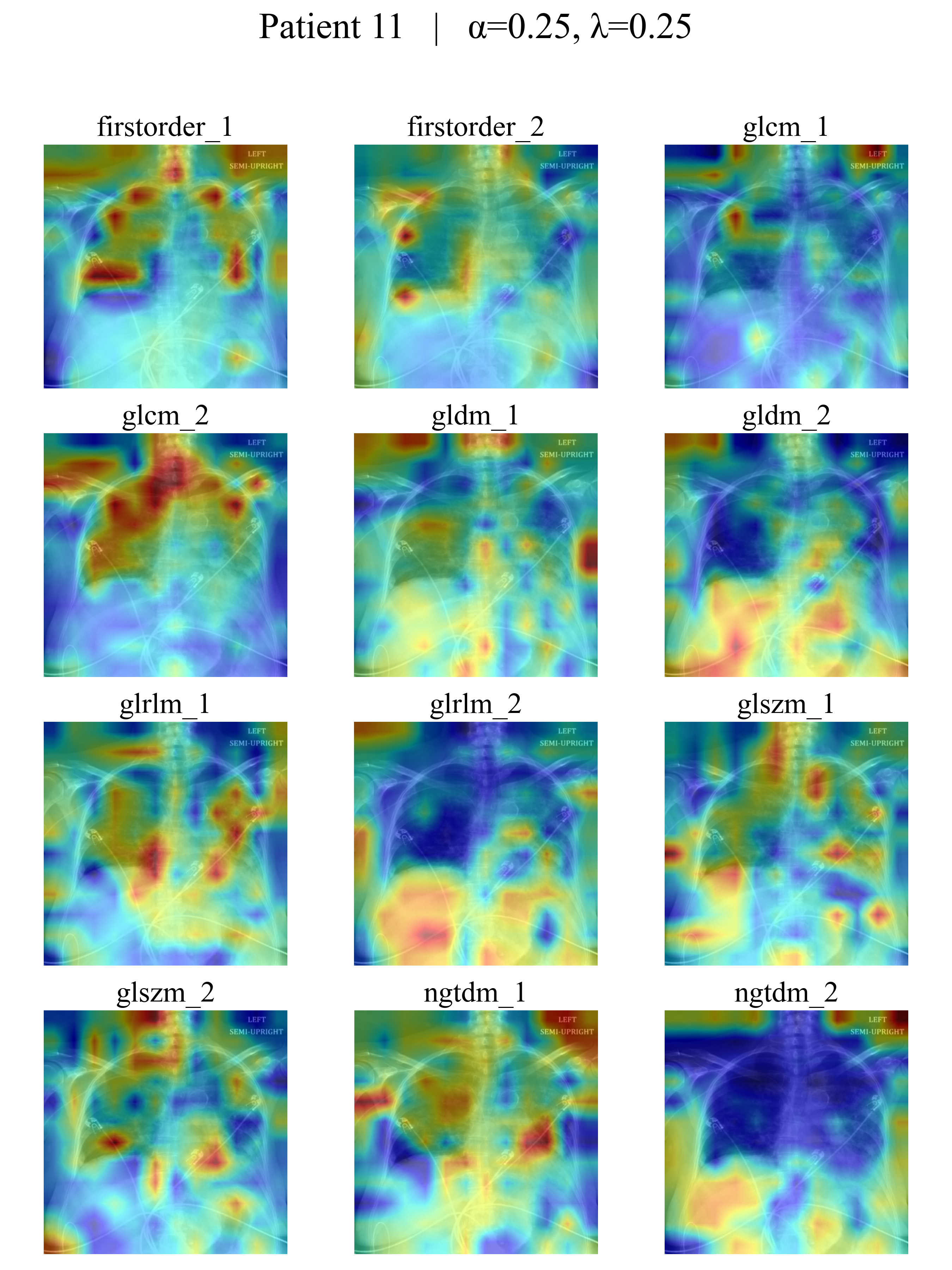}
\caption{CXR with highlighted regions guided by radiomics features (Patient 11).}
    \label{fig:fig_guide_case_2}
\end{figure}

As discussed, the advantage of the Radiomics-guided ViT is its added interpretability through feature-group–aware attention decomposition. To illustrate, see Figures~\ref{fig:fig_guide_case_1} and~\ref{fig:fig_guide_case_2}. They are the same two patients selected in Figure~\ref{fig:cxr_pix_case}. Now the Radiomics-guided ViT enables decomposition of the single CLS token attention into per–feature-group CLS token attentions, providing more interpretable insights into how different radiomic feature groups contribute to the model’s focus. Compared to Patient 11 in Figure~\ref{fig:cxr_pix_case}, we can see that the stronger attention in the right lung should be interpreted in terms of GLCM and First-order features.

Similarly, in equation~\eqref{eq:dual_rep}, a natural way to summarize feature importance is through the coefficients $\boldsymbol{\gamma}$, analogous to traditional regression settings. Thus we quantified the relative importance of each radiomics feature group by back-projecting the learned Cox coefficients (\(\boldsymbol{\gamma}\)) onto the original feature space. Specifically, for each group \(g\), the group-specific projection matrix and its corresponding Cox parameter vector were combined to obtain an \emph{effective} coefficient vector \(\boldsymbol{\gamma}_g\), representing how the model weighted the latent features derived from that group. The magnitude of these effective coefficients—summarized either by their mean absolute value (size-normalized) or by the L\(_2\) norm (total strength)—reflects the overall contribution of each group to the predicted hazard. This back-projection provides an interpretable measure of feature-group importance consistent with the linear structure of the Cox head.

\begin{figure}[ht]
\centering
\begin{minipage}{0.95\linewidth}
  \centering
  \includegraphics[width=\linewidth]{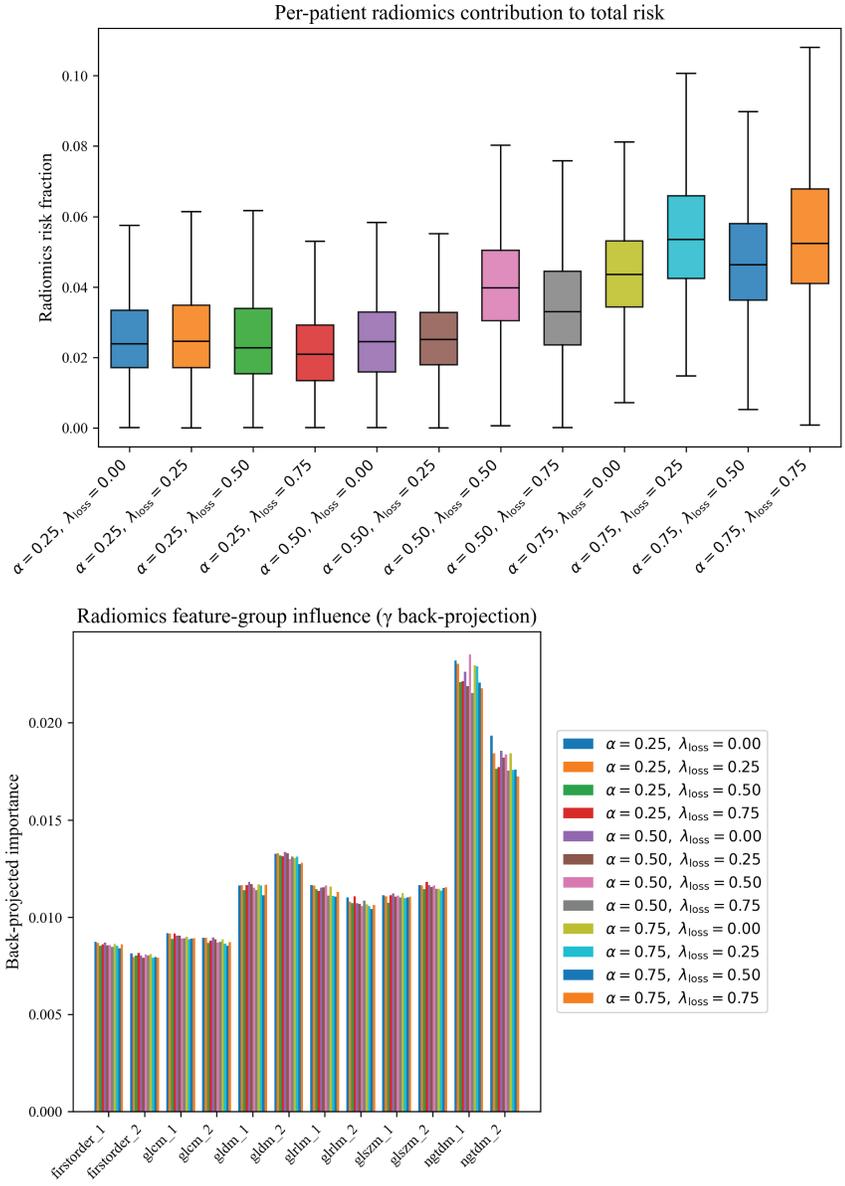}
  \vspace{2mm}
  \caption{Coefficient importance and risk contribution in the Radiomics-guided model.}
  \label{fig:fig_risk}
\end{minipage}
\end{figure}

Figure~\ref{fig:fig_risk} demonstrates how the hyperparameters $\alpha$ and $\lambda_{\text{loss}}$ change the risk estimation. While increasing either would increase the contribution of the radiomics branch, its average contribution remains below 5\% across all settings. This indicates that the ViT branch is still more influential in predicting survival. Meanwhile, the figure also illustrates the importance of feature groups. As shown, the dominant feature group is NGTDM. The estimation is consistent across all settings.

Lastly, we do not assume that the real data admit a latent structure in which only two underlying pattern vectors, $\boldsymbol{\mu}_1$ and $\boldsymbol{\mu}_2$, fully determine the hazard. Therefore, the dynamics conclusions derived in Section~\ref{sec:exp_dynamics} may not strictly hold in practice. Nevertheless, Figure~\ref{fig:fig_risk} provides a useful summary of feature-group importance.

\FloatBarrier

\section{Discussion}

This chapter bridged insights on ViT learning dynamics with practical survival modeling.
First, we extended the shallow-ViT framework of \cite{li2023theoretical} from classification to Cox proportional hazards loss.  
Second, through controlled experiments, we showed that pruning less informative tokens improves performance and interpretability in shallow ViTs, and that thresholding token-level attention changes can successfully recover label-relevant regions. Next, we established a baseline full ViT framework for both pixel- and radiomics-based inputs and extended it to a Radiomics-guided hybrid model, showing that radiomics-informed attention can improve interpretability without compromising predictive performance. The hybrid formulation, parameterized by the loss-weighting parameter $\lambda_{\text{loss}}$ and the modality-mixing parameter $\alpha$, offers flexible control over the contribution of each modality and can be fine-tuned through grid search to achieve balanced and robust performance across tasks. Overall, these findings unify theory, simulation, and real-data analysis—showing that ViT dynamics and pruning principles can extend to survival modeling, offering a path toward interpretable and computationally efficient transformer-based risk prediction in medical imaging.

We acknowledge that, in recent studies, deep learning methods and generative models have gained increasing popularity, including variational autoencoders (VAEs), generative adversarial networks (GANs), and diffusion models (e.g., \cite{kingma2013auto,goodfellow2014generative,ho2020denoising}). Though not discussed exhaustively here, we believe ViTs offer certain advantages over these alternative models. Taking VAE as an example, the distinction is primarily in how structure is captured rather than in adding modeling complexity. A transformer-based representation is designed to encode relationships across the entire input through attention, which allows it to preserve global dependencies and contextual interactions in a way that is directly aligned with downstream prediction. In contrast, a VAE-based representation is optimized to reconstruct the input under a generative objective, which may dilute features that are not essential for reconstruction but are important for the predictive target. As a result, the transformer representation tends to be more task-aligned, whereas the VAE representation is more distribution-aligned. The practical advantage is therefore not additional flexibility, but a closer match between the learned representation and the risk being minimized.

One limitation is that this study did not incorporate available longitudinal data, which include both clinical variables and multiple CXR images acquired as the disease progresses. Clinical information from electronic health records (EHR) is particularly valuable for modeling COVID-19 survival, while longitudinal imaging can capture dynamic patterns of disease evolution. In future work, we plan to integrate these longitudinal data into an illness–death survival framework to further investigate disease progression and outcomes.

\FloatBarrier


\bibliographystyle{plain}
\bibliography{ref}

\end{document}